\pdfoutput=1 
\documentclass[prb,singlecolumn,preprintnumbers,superscriptaddress,showpacs,preprint]{revtex4}

\usepackage{graphicx}
\usepackage{bm}
\usepackage{amsmath}
\usepackage{amssymb}
\usepackage[colorlinks,pdfstartview=FitH]{hyperref}
\hypersetup{linkcolor=blue,citecolor=blue,filecolor=black,urlcolor=blue}

\newcommand\be{\begin{eqnarray}}
\newcommand\ee{\end{eqnarray}}
\newcommand\p{{\bm p}}

\begin{document}

\author{Dmitri E. Kharzeev}
\affiliation{Department of Physics and Astronomy, Stony Brook University, 
Stony Brook, New York 11794-3800, USA}
\affiliation{Department of Physics, Brookhaven National Laboratory, 
Upton, NY 11973}
\author{Robert D. Pisarski}
\affiliation{Department of Physics, Brookhaven National Laboratory, 
Upton, NY 11973}
\affiliation{RIKEN-BNL Research Center, Brookhaven National Laboratory,
Upton, New York 11973-5000, USA}
\author{Ho-Ung Yee}
\affiliation{Physics Department, University of Illinois at Chicago, Chicago, 
Illinois 60607, USA}
\affiliation{RIKEN-BNL Research Center, Brookhaven National Laboratory,
Upton, New York 11973-5000, USA}

\title{Universality of plasmon excitations in Dirac semimetals}

\begin{abstract}

The recent experimental discovery of ${\rm Cd_3 As_2}$ 
\cite{Borisenko2014,Neupane2014} and ${\rm Na_3 Bi}$ \cite{Liu2014} 
Dirac semimetals enables the study of the properties
of chiral quasi-particles in three spatial dimensions.  
As demonstrated by photoemission \cite{Borisenko2014,Neupane2014,Liu2014}, 
Dirac semimetals are  
characterized by a linear dispersion relation for fermion quasi-particles, 
and thus represent three dimensional analogs of graphene.
While the distinctive behavior of chiral fermions 
({\it e.g.} Klein tunneling) is already evident in 
two dimensional graphene, 
the physics of chirality in three dimensions 
opens a number of new possibilities. 
In this paper we investigate the properties of the 
collective plasmon excitations in 
Dirac semimetals by using the methods of relativistic field theory. 
We find a strong and narrow plasmon excitation whose frequency is in the 
terahertz (THz) range which may be important for practical applications. 
The properties of the plasmon appear universal for all Dirac semimetals, 
due to the large degeneracy of the quasi-particles and the
small Fermi velocity, $v_F \ll c$. 
This universality is closely analogous to 
the phenomenon of ``dimensional transmutation'',
that is responsible for the emergence of dimensionful scales in 
relativistic field theories such as Quantum Chromodynamics, the modern
theory of nuclear physics.
\end{abstract}

\maketitle 

\section{Introduction}

The discovery of Dirac semimetals, which are chiral fermions 
in three spatial dimensions, opens the fascinating possibility 
of being able to study
the effects of quantum anomalies in condensed matter experiments. 
In particular, the presence of the chiral anomaly in 
$(3+1)$ dimensional theory 
should make it possible to observe ``Chiral Magnetic Effect (CME)'' --- 
a non-dissipative current induced by parallel electric and magnetic fields ---
in such systems; for a review, see \cite{Kharzeev2014}. 
The studies of magneto-transport in ${\rm Cd_3 As_2}$ 
have already begun \cite{Liang2014}.

The linear spectrum of quasi-particles also opens new possibilities for 
photonics/plasmonics. In graphene, which is two dimensional (2D),
the plasmon mode does not appear in the Random Phase Approximation (RPA)
\cite{RPA_graphene}.  A plasmon does arise after doping,
or the inclusion of electron-electron interactions, with a 
plasmon frequency that is in the terahertz (THz) range of frequency
\cite{Grigorenko2012}.  This range is important for 
diverse practical applications ranging from medical imaging to security. 

In this paper we investigate the properties of the collective 
plasmon excitation in three dimensional (3D) Dirac semimetals. 
Relative to 2D graphene, because of the extra spatial dimension 
a strong and narrow plasmon peak already appears in the Random Phase
Approximation.  At zero chemical potential and for a broad range 
in temperature, the plasmon frequency is 
approximately linear in $T$, and is in the THz range at room temperature.

Dirac semimetals are characterized by strong coupling and a
large fermion degeneracy, $N$.  We show that this leads to 
{\it universal} properties of the plasmon excitation: 
the plasmon spectrum does not depend on the value of the coupling constant
nor upon the degeneracy, $N$, of the Dirac point. 
The reason underlying this universality is the quantum scale anomaly of 
relativistic field theory, where it is known as ``dimensional transmutation". 
In Quantum Chromodynamics (QCD), this phenomenon is 
responsible for the masses of all strongly interacting particles, 
and thus for $\sim 95\%$ of the mass of the visible Universe. 

To compute the plasmon spectrum we need a method valid at strong coupling.
This is because for both 3D Dirac semimetals and for 2D graphene,
the role of the fine structure constant 
$\alpha_{em} = e^2/(4\pi \hbar c)$ is played by the 
effective coupling $\alpha = e^2/(4\pi \hbar v)$, where 
$v \ll c$ is the Fermi velocity. 
The Fermi velocity in ${\rm Cd_3 As_2}$ was 
experimentally determined\cite{Liang2014} to be 
$v \simeq 9.3 \times 10^5\ {\rm m/s} \simeq 1/300\ c$, 
close to the value in graphene. 
Because of this, the effective coupling constant $\alpha \simeq 2.2$ is 
very large.  This is comparable to the value of the strong coupling constant
in the Quark-Gluon Plasma, near the deconfining transition in QCD.

Generally, the photon propagator cannot be computed perturbatively in strong
coupling.  However, there is an alternate expansion possible.
The degeneracy factor of fermion quasi-particles is large:
due to the degeneracy in the electron spin and double valleys, $N = 4$ for both
3D Dirac semimetals and for graphene.
We can then use a large $N$ expansion to compute the photon propagator
to leading order in $1/N$.  At nonzero temperature and density,
the result for the photon propagator is similar to that obtained
in the Hard Thermal Loop (HTL) approach to the Quark-Gluon Plasma
\cite{Pisarski1989,Braaten1990}.  The HTL approximation is used, {\it e.g.},
to compute the rate of electromagnetic radiation 
from the Quark-Gluon Plasma \cite{braaten1990prod}. 
In this paper we employ similar methods
for evaluating the plasmon spectrum and damping rate in Dirac semimetals.

When the number of fermions species $N$ is large, 
the photon dynamics is dominated by dressing the photon 
with the one loop fermion diagrams.
As a result, the photon propagator is suppressed by $1/N$, and
photon-mediated interactions are suppressed by $1/N$, so the fermion
dynamics are those of a free theory.
As long as $N$ is sufficiently large, this remains true even at strong
coupling \cite{son2007quantum}. 
Further, 
we can neglect the scale dependence of the Fermi velocity, as that
originates from loop corrections to the fermion propagator.
Indeed, for graphene the suppression of the dependence of the
velocity scale with $1/N$ is manifest  \cite{aleiner2007spontaneous}. 
 
In the one loop approximation at large $N$, the longitudinal
Coulomb and the transverse sectors of the plasmon decouple from one another.
In the following we focus on the Coulomb sector.  
(We note,
however, that in the static
limit in which we compute, the transverse and Coulomb plasmons are
degenerate.)
From the above discussion, the effective coupling in the Coulomb sector is
\be
\lambda(\Lambda_c)\equiv {N e^2(\Lambda_c)\over v}\,.
\ee
We emphasize here the dependence on the physical UV lattice cutoff 
$\Lambda_c$ at which the observed value of the coupling  is defined:
\be
e^2(\Lambda_c) \simeq {1\over 137}\times (4\pi)\approx 0.1\,.
\ee

It is well known that in a gauge theory with massless 
fermions there is no intrinsic notion of the coupling constant:
the coupling constant changes, or ``runs'', as the length scale at which
it is probed changes.  
Hence one can trade the value of the coupling constant for the dimensionful
scale at which it is defined, which is known as ``dimensional transmutation''.
For our purposes we can define this scale as that where the coupling
blows up, at the Landau pole $\Lambda_L$.
The  physical observables then depend only upon 
the ratio of an external scale, $Q$, at which the coupling is measured
to $\Lambda_L$. 
At one loop order, the coupling $\lambda(Q)$ at a scale $Q$ is given by 
\be\label{run_couple}
\lambda(Q)={\lambda(\Lambda_c)\over 
1-\frac{\lambda(\Lambda_c)}{12\pi^2}\log\left(
Q^2\over \Lambda_c^2\right)} =
{12\pi^2\over \log\left(\Lambda_L^2\over Q^2 \right)}\, .
\ee
The first equality in Eq. (\ref{run_couple}) contains a Landau pole at
$\Lambda_L\equiv\Lambda_c\cdot \exp(6\pi^2/\lambda(\Lambda_c))$,
which is where the coupling constant diverges. 
We can then rewrite this as the second equality in Eq. (\ref{run_couple}),
which shows that the coupling is a function solely of the ratio 
$Q/\Lambda_L$.  That is,
$\Lambda_c$ and $\lambda(\Lambda_c)$ are 
transmuted to a single scale $\Lambda_L$, 
which is the only dimensionful parameter of the theory. 
This means that at nonzero temperature $T$ and chemical potential $\mu$, 
any observable in the photon sector is of the form
\be
T^\Delta f\left({T\over\Lambda_L}, {\mu\over T}\right)\, .
\label{full}
\ee 
Here $\Delta$ is the mass dimension of the observable; in this
paper it is the plasmon frequency, with $\Delta=1$.
The function $f(x, y)$ depends upon the observable in question,
but is otherwise universal: 
all of the dependence on $\Lambda_c$ and 
$\lambda(\Lambda_c)$ is included in the single parameter, $\Lambda_L$. 
It is worth emphasizing that neither $N$ or $v$
appears in the function $f(x, y)$.  This is most powerful,
as it is then possible to find $f(x,y)$ with ease in the one-loop approximation
valid at large $N$.  
In this paper we compute the universal function $f(x, y)$ for
the plasmon frequency at zero spatial momentum.
It is worth emphasizing that the ``vacuum'' contribution to the one loop
diagram, from zero temperature and density, plays a crucial role in
realizing this universality.  (This is not captured by the Hard Thermal Loop
limit, which neglects the vacuum contribution.)

The limit of strong coupling is defined as follows.
Given the physical lattice cutoff, $\Lambda_c$, 
with a fixed $e^2(\Lambda_c)\approx 0.1$, 
a large value of $N/v$ can give a large value of $\lambda(\Lambda_c)$. 
Thus, in the strong coupling limit, $\Lambda_L =
{\rm e}^{6 \pi^2/\lambda(\Lambda_c)} \Lambda_c \approx \Lambda_c$,
any observable in the photon sector
at nonzero $T$ and $\mu$ becomes
\be
T^\Delta f\left({T\over \Lambda_c}, {\mu\over T} \right)\,,\label{strong}
\ee
with the {\it same} function $f(x, y)$.
That is, the result is independent of the values of 
$N$, $v$, or $e^2(\Lambda_c)\approx 0.1$.  We call this a universality
of strong coupling.

For $N=4$ and $1/v=300$, we have
\be
\lambda(\Lambda_c)/6\pi^2\approx 2\,,
\ee
so that
\be\label{num_lam}
\Lambda_L\approx 1.65\,\, \Lambda_c\,.
\ee
In spite of the uncertainty in the value of $\Lambda_c$, 
it is thus reasonable to assume that the strong coupling limit, 
and so Eq. (\ref{strong}), are justified.

We now turn to a summary of the details of the computation of the plasmon
frequency.  The plasmon arises from the singularity in the 
longitudinal component of the retarded photon propagator.
In Coulomb gauge, $\vec\nabla\cdot\vec A=0$, this propagator is 
\be
\Pi^{00}_{ R}(p)\equiv \langle A^0(p) A^0(-p)\rangle=
{i\over |\p|^2-\Pi^L_R(p)}\,,\quad \Pi^L_R(p)\equiv \langle J^0(p) 
J^0(-p)\rangle_R\,.
\label{pi00}
\ee
The one-loop expression for the longitudinal retarded 
self-energy $\Pi^L_R(p)$ consists of two parts:
the first in vacuum, at $T = \mu = 0$,
and the second from $T, \mu \neq 0$. 
The explicit computation is presented in the Methods section.  This
shows that the solution for the plasmon frequency takes the form
\be
\omega_{pl}(T) =T \,\,f(T/\Lambda_L, \mu/T)\,,\label{univ}
\ee
where the function $f(x,y)$ is universal,
independent of the values of
the coupling constant $e^2(\Lambda_c)$, degeneracy $N$, and 
the Fermi velocity $v$.

\begin{figure}
  \centering
  \includegraphics[width=80mm]{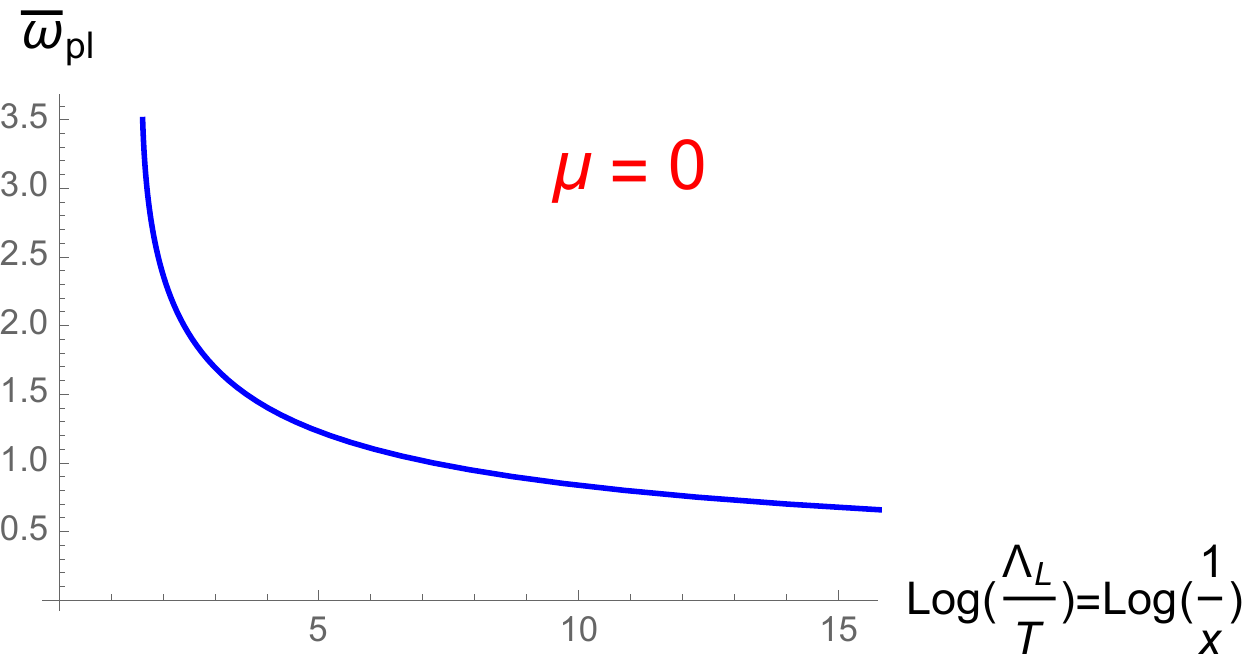}  \includegraphics[width=80mm]{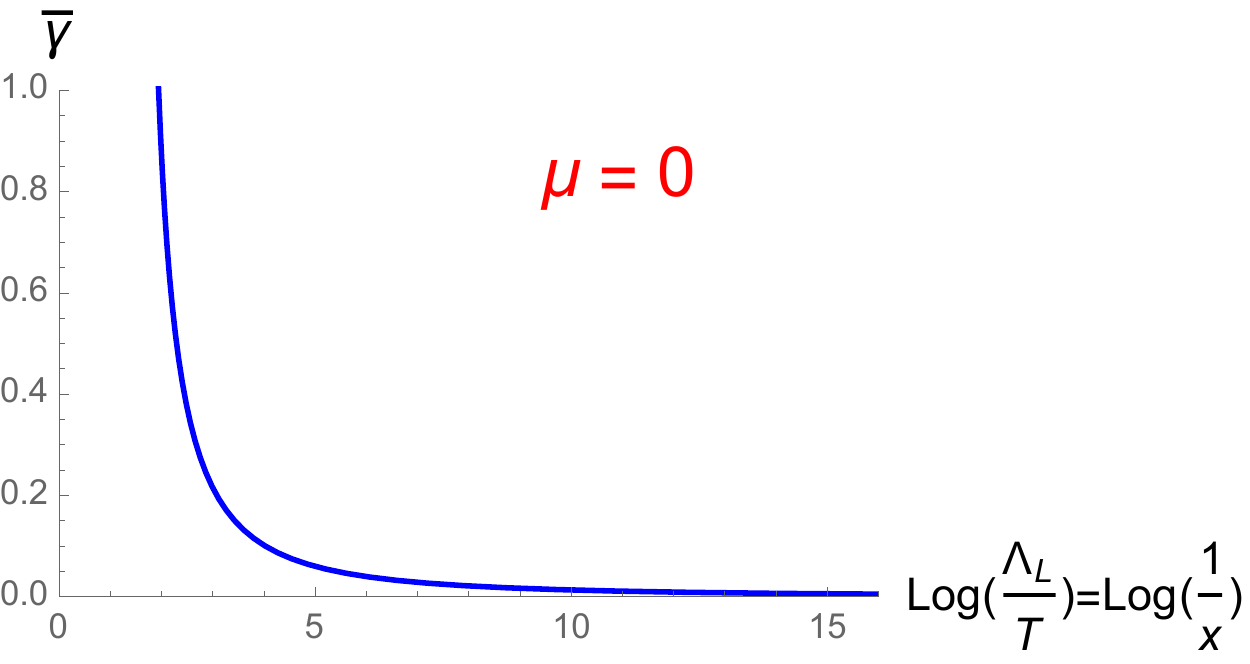}
  \caption{The real (left) and imaginary (right) parts of 
the plasmon energy, divided by the temperature, 
$f(x) \equiv \omega_{pl}/T - i\gamma/T$,
as a function of $\log(\Lambda_L/T) = \log(1/x)$.}
  \label{fig1}
\end{figure}

For small $x \equiv T/\Lambda_L$ the function $f(x,y)$ 
can be found analytically which agrees with the HTL  
with running coupling constant. The result is 
\be
f(x,y)\approx \sqrt{2 \left({\pi^2\over 3} + y^2\right)\over\log(1/x)}
\,,\quad x\ll 1\,.\label{xl1}
\ee
For a general $x$ and $y$, $f(x,y)$ is complex valued; 
we have evaluated it numerically as described in Methods. 
The resulting real and imaginary parts of the plasmon energy,
normalized to the temperature $T$, 
are presented in Fig. (\ref{fig1}) as a function of 
$\log(\Lambda_L/T) = \log(1/x)$. 

\begin{figure}
  \centering
  \includegraphics[width=100mm]{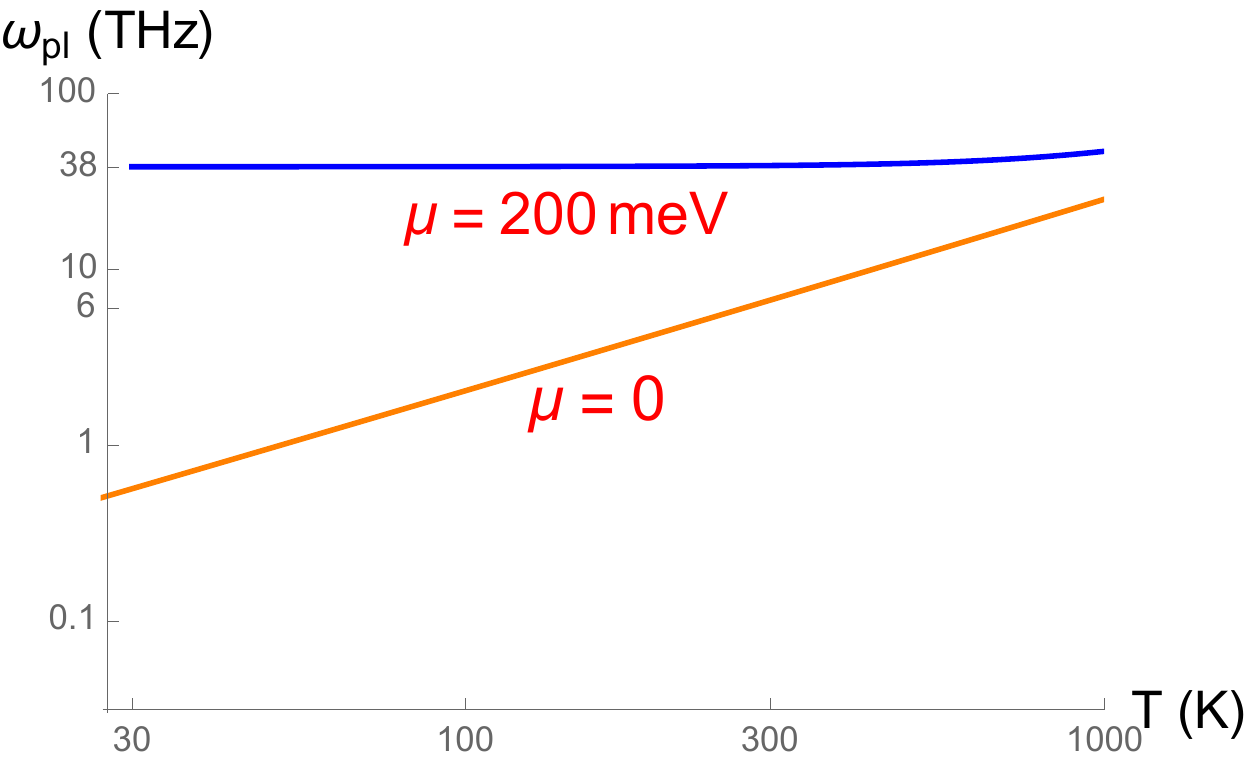}  
\caption{The plasmon frequency $\omega_{pl}$ as a function of temperature $T$.}
  \label{fig2}
\end{figure}

In Fig. (\ref{fig2}) we present the plasmon frequency in 
physical units of terahertz (THz) as a function of 
temperature at zero chemical potential and at a
chemical potential of $\mu = 200$ meV, which is 
characteristic for  ${\rm Cd_3 As_2}$\cite{Borisenko2014,Neupane2014}. 
For the case of zero chemical potential, 
we see that by changing the temperature, 
the plasmon frequency can be tuned from the radio wave to the 
near infrared range of the spectrum. 
In this entire frequency range, the damping of the plasmon is weak, 
with $\gamma/\omega_{pl} < 0.05$, so the plasmon peak is very narrow. 

Let us first estimate numerically the magnitude of the plasmon frequency that we have derived. 
The UV cutoff in the energy spectrum of quasiparticles indicated by the ARPES measurements in ${\rm Cd_3 As_2}$
is $\Lambda_c \simeq 5\ {\rm eV} \simeq 5.8 \times 10^4\ {\rm K}$. 
For the dimensionful scale $\Lambda_L$ we thus get  
 $\Lambda_L \simeq 1.65\ \Lambda_c \simeq 10^5$ K, see Eq. (\ref{num_lam}).
 For the room temperature of $T \simeq 300$ K we get 
$\log(\Lambda_L/T) \simeq 5.7$. Fig. (\ref{fig1}) 
then yields the plasmon frequency of $\omega_{pl} \simeq T \simeq 6$ THz $\simeq 0.5\ \rm{mm}^{-1}$. We have thus found that for room temperature  ${\rm Cd_3 As_2}$  possesses the plasmon in the terahertz frequency range, which may have important applications for THz imaging. It is known that ${\rm Cd_3 As_2}$ undergoes a phase change at the temperature of $T \simeq 888$ K 
\cite{Hiscocks69}. 
For this temperature, we get  $\log(\Lambda_L/T) \simeq 4.7$, and from 
Fig. (\ref{fig1}) the plasmon frequency is still $\omega_{pl} \simeq T$, which at this higher temperature yields a higher frequency $\omega_{pl} \simeq 18$ THz. At a low temperature of $T = 3$K, we get $\log(\Lambda_L/T) \simeq 10.4$,
and from Fig. (\ref{fig1}) the plasmon frequency is $\omega_{pl} \simeq T$ which yields a low frequency of $\omega_{pl} \simeq 60$ GHz $\simeq 1\ \rm{cm}^{-1}$ which is in the radio frequency range. 

In summary, plasmons in Dirac semimetals provide a link between the
quantum dynamics of relativistic field theories and photonics.
Depending on the chemical potential, which can be controlled by doping, 
Dirac semimetals can be used as sensors or emitters of 
electromagnetic radiation in a broad frequency range, between 
radio waves, $\sim 100$ GHz, and near infrared, $50$ THz.

\newpage

\section{Methods}

Consider the longitudinal component of the
retarded photon propagator in Coulomb gauge, $\vec\nabla\cdot\vec A=0$,
as given in Eq. (\ref{pi00}).  
%
The contribution in vacuum, at $T = \mu = 0$, is
%
\be
\Pi^{\rm vac}_R(p)={Ne^2(\Lambda_c)|\p|^2\over 12\pi^2 v}
\log\left(-p^2\over\Lambda_c^2\right)\Big |_{p^0\to p^0+i\epsilon}\, .
\ee 
The square of the four momentum is 
$p^2 = (p^0)^2 - v^2 |\p|^2$. 
$\Lambda_c\approx 5 \,\,{\rm eV}\approx 10^5\,\,K$ is the ultraviolet 
cutoff in the energy spectrum of chiral quasiparticles, with 
the value  indicated by the ARPES data 
on ${\rm Cd_3 As_2}$\cite{Borisenko2014,Neupane2014}.

The second part is the contribution from $T$, $\mu \neq 0$.  After
rescaling the spatial momentum integration variable from ${\bm k}$
to ${\bm k}/v$, this is given by
\be
\Pi^{\rm th}_R(p)={Ne^2(\Lambda_c)\over2\pi^2 v^3}
\int_0^\infty dk\,\,k^2 {\cal N}(k) I(p,k)\,,
\ee
where
\be
I(p,k)&=&\int^{1}_{-1}dx\,\,
\Bigg({2k+p^0+v|\p| x\over (p^0)^2-v^2|\p|^2+2p^0 k-2v|\p|kx+i\epsilon(k+p^0)}
\nonumber\\
&&-{2k-p^0-v|\p| x\over -(p^0)^2+v^2|\p|^2+2p^0 k-2v|\p|kx
+i\epsilon(k-p^0)}\Bigg)\, ,
\ee
where ${\cal N}(k)=(e^{(k-\mu)/T}+1)^{-1} + (e^{(k+\mu)/T}+1)^{-1}$ 
is the sum of the Fermi-Dirac statistical 
distribution functions for particles and antiparticles (holes). 
In principle, the integration over $k$ has ultraviolet cutoff
from the lattice.  However, as ${\cal N}(k)$ is Boltzmann suppressed
at large momenta, we can extend the integration range to infinity,
up to exponentially small corrections.
The $x$ integration can be explicitly computed by logarithms. 
%
In the limit of small spatial momenta, $\p \to 0$, 
this becomes
\be
I(p,k)\to {4\over 3}{k v^2 \, \p^2\over(p^0)^2
\left(k+{p^0\over 2}+i\epsilon\right)\left(k-{p^0\over 2}-i\epsilon\right)}
+{\cal O}((\p^2)^2)\,.
\ee
In the limit of $\p \rightarrow 0$, then, the equation for the pole
in the longitudinal propagator becomes
\be
&&(p^0)^2\left(1-{N e^2(\Lambda_c)\over 12\pi^2 v}\log\left(-(p^0+i\epsilon)^2\over\Lambda_c^2\right)\right)\nonumber\\&-&{4N e^2(\Lambda_c)\over 3\pi^2 v}\int^\infty_0 dk\,\,{\cal N}(k) {k^3\over \left(k+{p^0\over 2}+i\epsilon\right)\left(k-{p^0\over 2}-i\epsilon\right)}=0\,.
\ee
As discussed previously, $\lambda = N e^2/v$ appears naturally.
Also, the vacuum term reflects the running of the coupling constant.
The above equation can be rewritten as
\be
(p^0)^2\log\left(-(p^0+i\epsilon)^2\over\Lambda_L^2\right)+16\int^\infty_0 dk\,\,{\cal N}(k) {k^3\over \left(k+{p^0\over 2}+i\epsilon\right)\left(k-{p^0\over 2}-i\epsilon\right)}=0\, . 
\label{eq}
\ee
In this expression, all other parameters disappear, and are
replaced by the single scale $\Lambda_L \approx 1.65 \Lambda_c$.
From the above equation, it is clear that the solution for the 
plasmon frequency takes the universal form
\be
\omega_{pl}(T) =T \,\,f(T/\Lambda_L, \mu/T)\, .
\label{univ}
\ee
We determine the function $f(x,y)$ shortly. 
This is an explicit demonstration of the universality which we argued above.
To see the emergence of the form given by Eq. (\ref{univ}) more clearly, 
we change the integration variable $k\to {\bar k} = k/T$, 
introduce ${\bar \mu} = \mu/T$, and write the Eq. (\ref{eq}) as 
\be\nonumber
(\bar p^0)^2\log\left(-(\bar p^0+i\epsilon)^2\cdot 
{T^2\over\Lambda_L^2}\right)
\ee
\be
+ 8\int_0^\infty d{\bar k}\, {\bar k}^3\,\left({1\over e^{{\bar k} - {\bar \mu}}+1} + {1\over e^{{\bar k} + {\bar \mu}}+1}\right) {1\over \left({\bar k}+{\bar p^0\over 2}+i\epsilon\right)\left({\bar k}-{\bar p^0\over 2}-i\epsilon\right)}=0\,,
\ee
where $\bar p^0=p^0/T$.

The above equation is our starting point to find the universal function
$f(x, y)$ for the plasmon frequency.
We emphasize that it is the  full expression to one loop order in 
the static limit, $\p\to 0$.

Before presenting our numerical solution for $f(x, y)$, it is 
instructive to study its behavior for $x\ll 1$ and $x\gg 1$ at fixed $y$.
For $x\ll 1$, one can replace the logarithm in the equation by $2\log x$;
we can then neglect $\bar p^0\ll 1$ in the $k$ integration, so that
the equation becomes 
\be
2(\bar p^0)^2 \log x+8\int^\infty_0 d{\bar k}\,{\bar k} 
\left({1\over e^{{\bar k} - {\bar \mu}}+1} + {1\over e^{{\bar k} + {\bar \mu}}+1}\right)  =2(\bar p^0)^2 \log x+{4\pi^2\over 3} + 4 {\bar \mu}^2=0\,,
\ee
which gives
\be
f(x,y)\approx \sqrt{2 \left({\pi^2\over 3} + y^2\right)\over\log(1/x)}\,,\quad x\ll 1\,.\label{xl1}
\ee
For $x\gg 1$, again we replace the logarithm by $2\log x$, 
as then $\bar p^0\ll 1$, and the $k$ integration is the same as before. 
There is no consistent real solution in this case since $\log x>0$. 
Therefore, the real part of $f(x, y)$ ceases to exist for $x>x_c$ 
with some critical number $x_c$, equivalently for a
temperature $T>T_c=x_c \Lambda_L$. That is, plasmons do not exist in this
case.

For a general $x$ and $y$, $f(x,y)$ is complex valued. 
For $x\ll 1$ it is easy to find that its imaginary part is 
sub-leading compared to the real part of Eq. (\ref{xl1}).
As $x$ increases, the imaginary part becomes larger, and eventually dominates
for $x\gg 1$. 
In our numerical analysis we find that the imaginary part is
always small relative to the real part, so that it can be determined in
a linear approximation.  

The result of Eq. (\ref{xl1}) can also be obtained 
by using the HTL approximation with a running coupling constant. 
The HTL result for the plasmon frequency is
\be\label{pl_htl}
\omega_{pl} = \sqrt{\frac{N}{v}}\  e \frac{T}{3} = \frac{\sqrt{\lambda}}{3}\ T .
\ee 
If we use the running coupling of Eq. (\ref{run_couple}) in Eq. (\ref{pl_htl}),
\be
\lambda (T) = \frac{12 \pi^2}{\log \left(\frac{\Lambda_L^2}{T^2}\right)} ,
\ee
then we reproduce the expressions of Eqs. (\ref{univ}) and (\ref{xl1}).

Writing 
\be
f(x)=\bar p^0=\frac{\omega_{pl}}{T}-i \frac{\gamma}{T} \equiv 
{\bar\omega}_{pl}-i{\bar\gamma} \; ,
\ee
the equation for the real part of ${\bar\omega}_{pl}>0$ is
\be
{\bar \omega}_{pl}^2\left(
\log x+\log \bar{\omega_{pl}}\right)
+4 {\cal P}\int^\infty_0 d{\bar k} \,   
\left({1\over e^{{\bar k} - {\bar \mu}}+1} + {1\over e^{{\bar k} + {\bar \mu}}+1}\right) 
{{\bar k}^3\over ({\bar k}+{\bar\omega}_{pl}/2)({\bar k}-\bar{\omega}_{pl}/2)}
=0\,,
\ee
where $\cal P$ denotes a principal value integration. 
With ${\bar{\omega}}_{pl}$ found, 
in linear approximation the damping rate $\gamma$ is
\be
\bar{\gamma}=-{\pi\over 4}{{\bar\omega}_{pl}\over 
(\log x+\log{\bar\omega}_{pl})}
\left(1 - \frac{1}{e^{({{\bar\omega}_{pl}/2}-{\bar\mu})}+1} 
- \frac{1}{e^{({{\bar\omega}_{pl}/2}+{\bar\mu})}+1}\right) .
\ee
In Figures (\ref{fig1}) and (\ref{fig2}) 
we show the numerical plots of $f(x,y)$ at fixed $\mu=y T$. 
The real part of $f(x,y)$ ($\bar{\omega}_{pl}$) 
ceases to exist for $x>x_c\approx 0.13$
with the maximum value around 3.5. 
In the most relevant region of $x$ with $\log(1/x)\gtrsim 5$, 
the imaginary part (the damping rate $\bar\gamma = \gamma/T$) 
is indeed less than $5\%$ of the real part, so the plasmons are 
narrow and so well defined excitations.

\section{acknowledgements}
We thank D. Son and M. Stephanov
for discussions.  This work was supported in part by the U.S.
Department of Energy under Contracts No. DE-FG-
88ER40388 and DE-AC02-98CH10886.  

\bibliographystyle{my-refs}

\bibliography{dirac.bib}

\end{document}